\documentclass[12pt]{article}
\usepackage{pic03}
\usepackage{hyperref}
\usepackage{url}
\usepackage{graphicx}

\begin{document}

\title{\bf  MEASUREMENT OF THE ANGLE $\phi_1$($\beta$) AND $B\bar{B}$ 
MIXING (RECENT RESULTS FROM BaBar AND Belle)
}

\author{Kazuo Abe        \\
{\em KEK, Tsukuba, Japan 305-0801}}
\maketitle

%
%
%
%
%
%
\vspace{4.5cm}
%

\baselineskip=14.5pt
\begin{abstract}
Recent results from BaBar and Belle experiments on $B \bar B$ mixing and
 $\sin 2\phi_1$ are presented. Accuracy of $\Delta m_d$ measurements has reached
 1.2\%. Higher order effects within the Standard Model or possible new
 physics effect that might appear in the $B \bar B$ mixing
 through non-zero $\Delta \Gamma/\Gamma$, $CP$ violation, or $CPT$
 violation have been explored.  The BaBar and Belle results on $\sin 2\phi_1$ from 
the $b \to c \bar{c} s$ modes are in good agreement with each other and 
 a combined result with an accuracy of 8\% is in good agreement with a
 global CKM fit. A simple average of the $\sin 2\phi_1$ values that were 
 measured in the penguin-loop dominated decay modes, $\phi K_S$, 
$\eta^{\prime} K_S$ , and $K^+ K^- K_S$, 
shows about 2.5$\sigma$ deviation from the Standard Model.
\end{abstract}
\newpage

\baselineskip=17pt

\section{ $e^+ e^- \to \Upsilon(4S) \to B \bar{B}$}
\begin{figure}[htbp]
  \centerline{\hbox{ \hspace{0.2cm}
    \includegraphics[width=6.5cm]{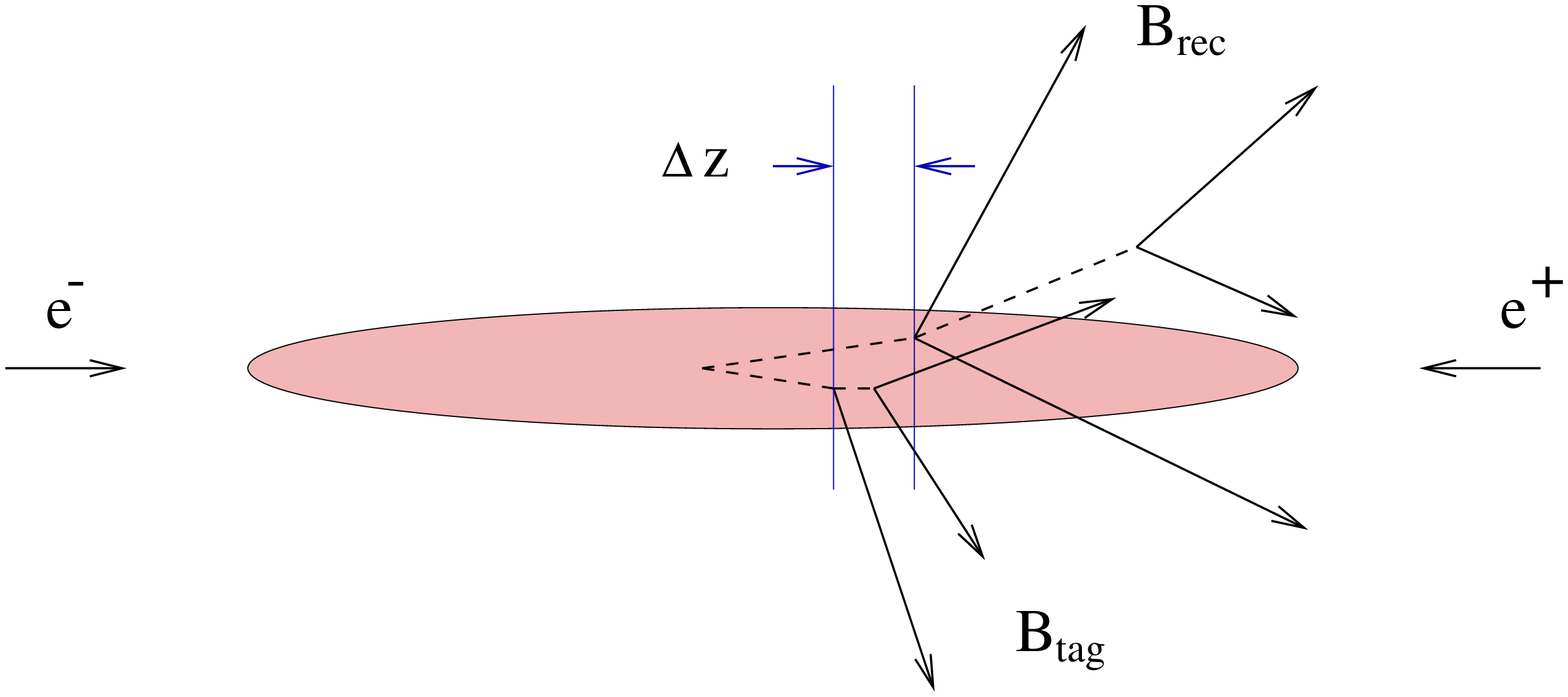}
    }
  }
 \caption{\it
      Schematical drawing of $e^+ e^- \to \Upsilon(\mathrm{4S}) \to B \bar{B}$ 
      process at PEP-II and KEKB.
    \label{IR} }
\end{figure}

A scheme of producing $\Upsilon (\mathrm{4S})$ in an asymmetric-energy 
$e^+ e^-$ collisin, that is used at PEP-II and KEKB, enables 
separation of the decay verteces of the two $B$ mesons. 
PEP-II operates at $9~\mathrm{GeV}~e^- \times 3.1~\mathrm{GeV}~e^+$ 
corresponding to $\Delta z \simeq 260\mu \mathrm{m}$, while 
KEKB operates at $8~\mathrm{GeV}~e^- \times 3.5~\mathrm{GeV}~e^+$
corresponding to $\Delta z \simeq 200\mu \mathrm{m}$. 
Since the size of interaction region in the $z$ direction is much larger than
these $\Delta z$ ($\sim 7\mathrm{mm}$ at KEKB), the reference of the
proper time must be the decay point of the other $B$ (See Fig.~\ref{IR}).    
Conservation of charge-conjugation in the $\Upsilon
(\mathrm{4S}) \to B^0 \bar{B^0}$ decay forces  
the time structure of $B \bar B$ system to stay 
as $\psi(t)=|B^0>|\bar{B^0}>-|\bar{B^0}>|B^0>$ 
at any $t$ until one $B$ meson decays. This feature is used to determine
the flavor of the reconstructed $B$ at $\Delta t = 0$. 

\section{ $B \bar{B}$ Mixing}
\begin{figure}[tp]
  \centerline{\hbox{ \hspace{0.2cm}
    \includegraphics[width=3.5cm]{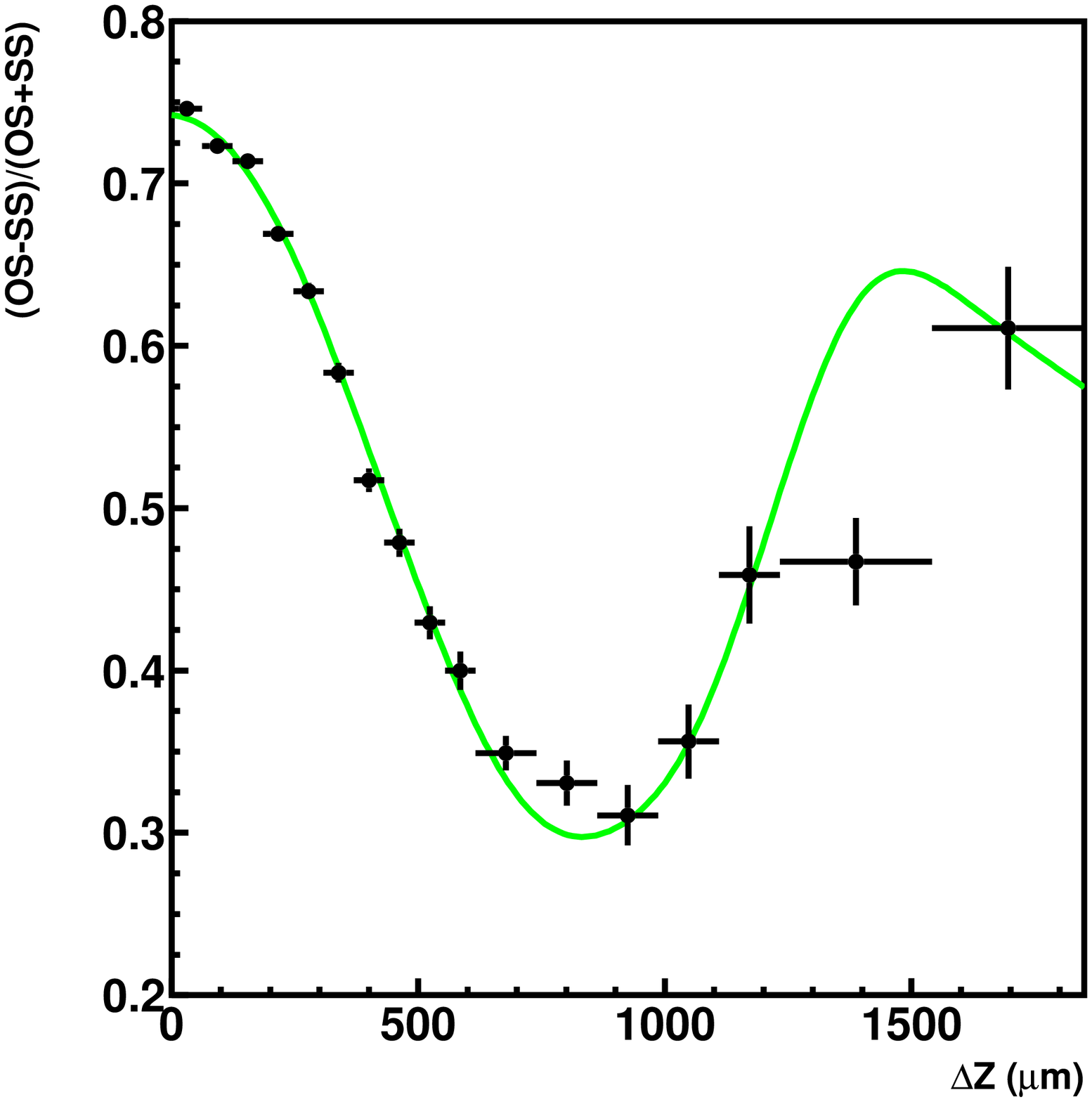}
    \includegraphics[width=3.5cm]{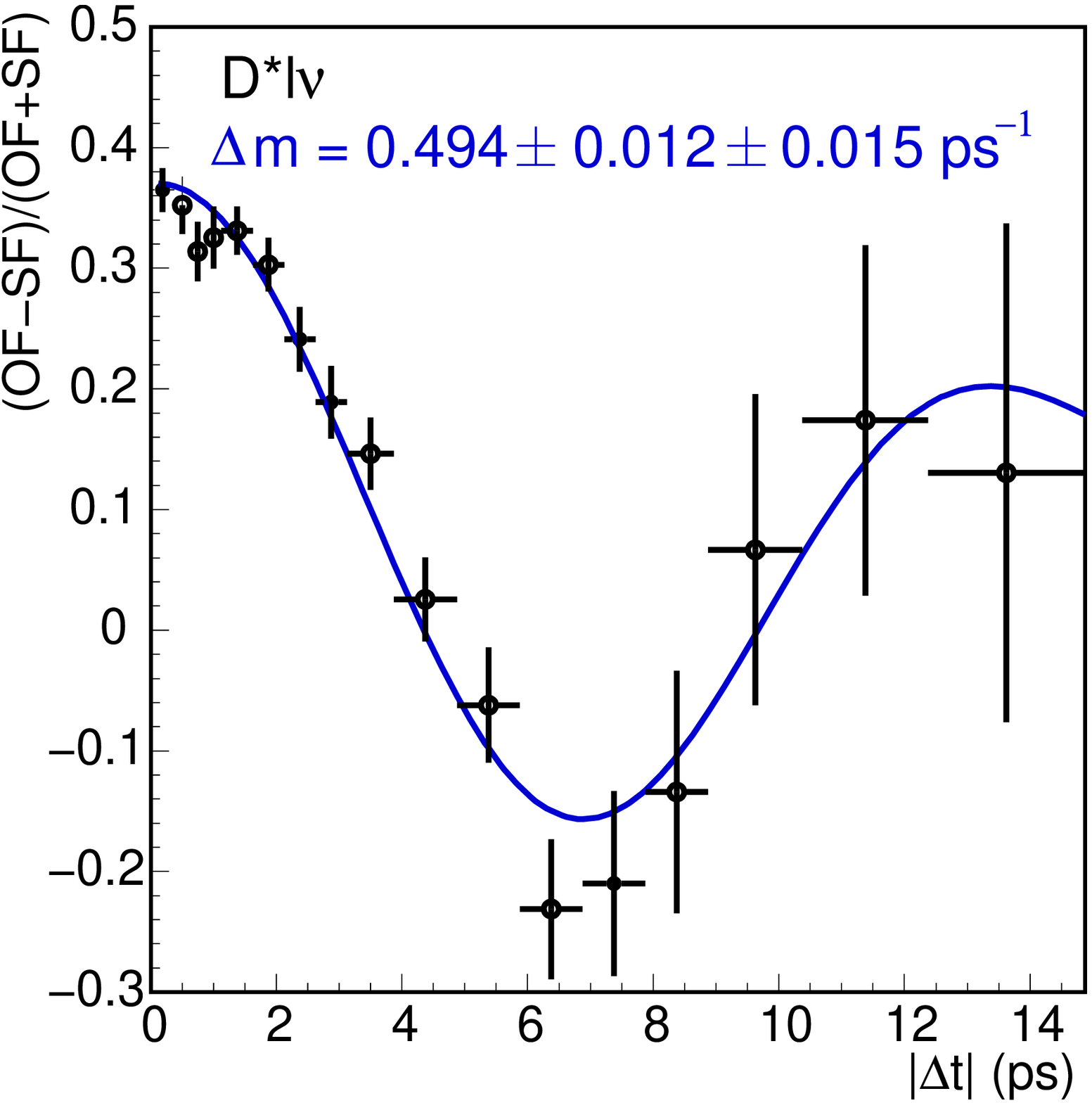}
   \includegraphics[width=3.5cm]{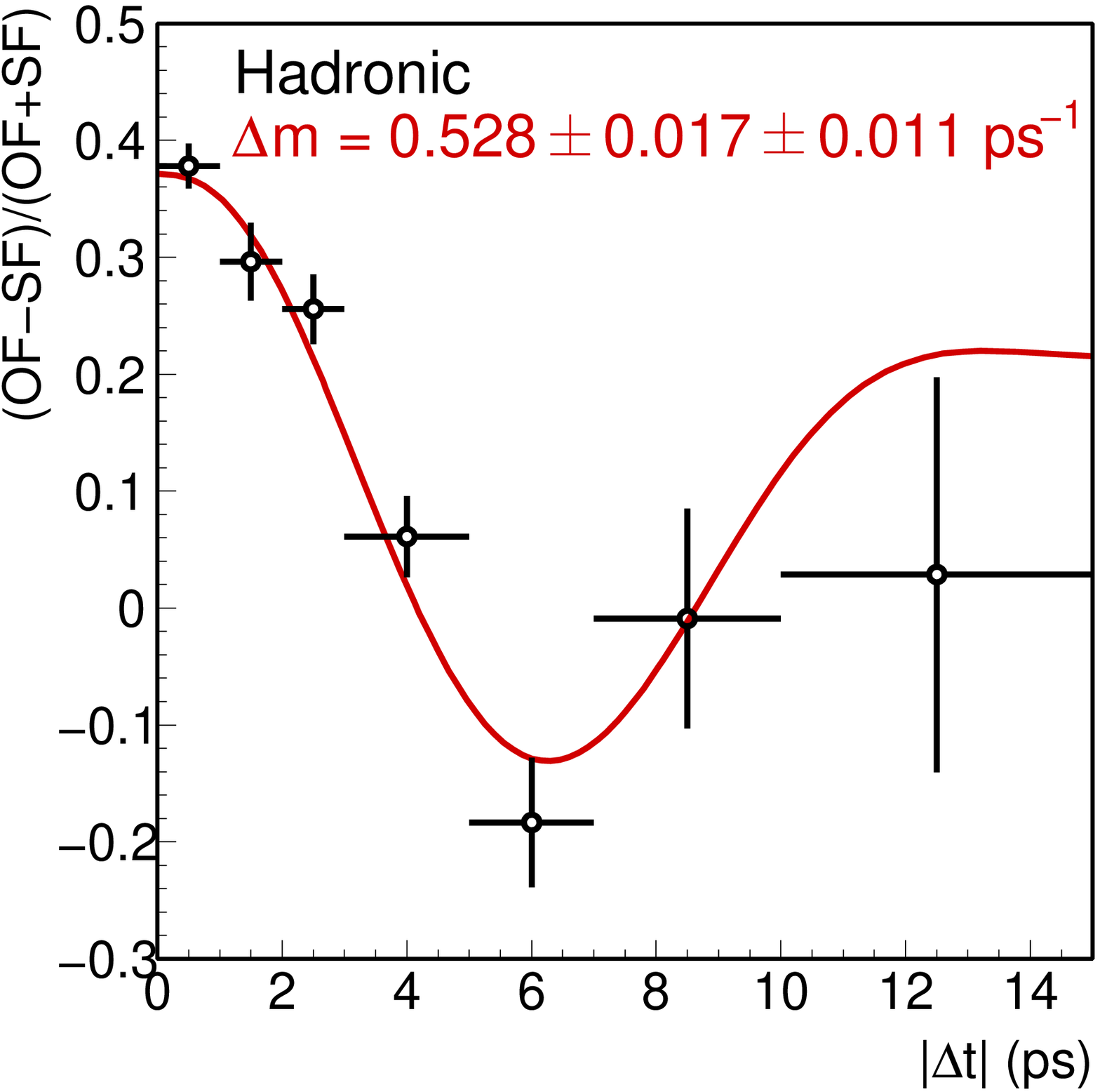}
  \includegraphics[width=3.5cm]{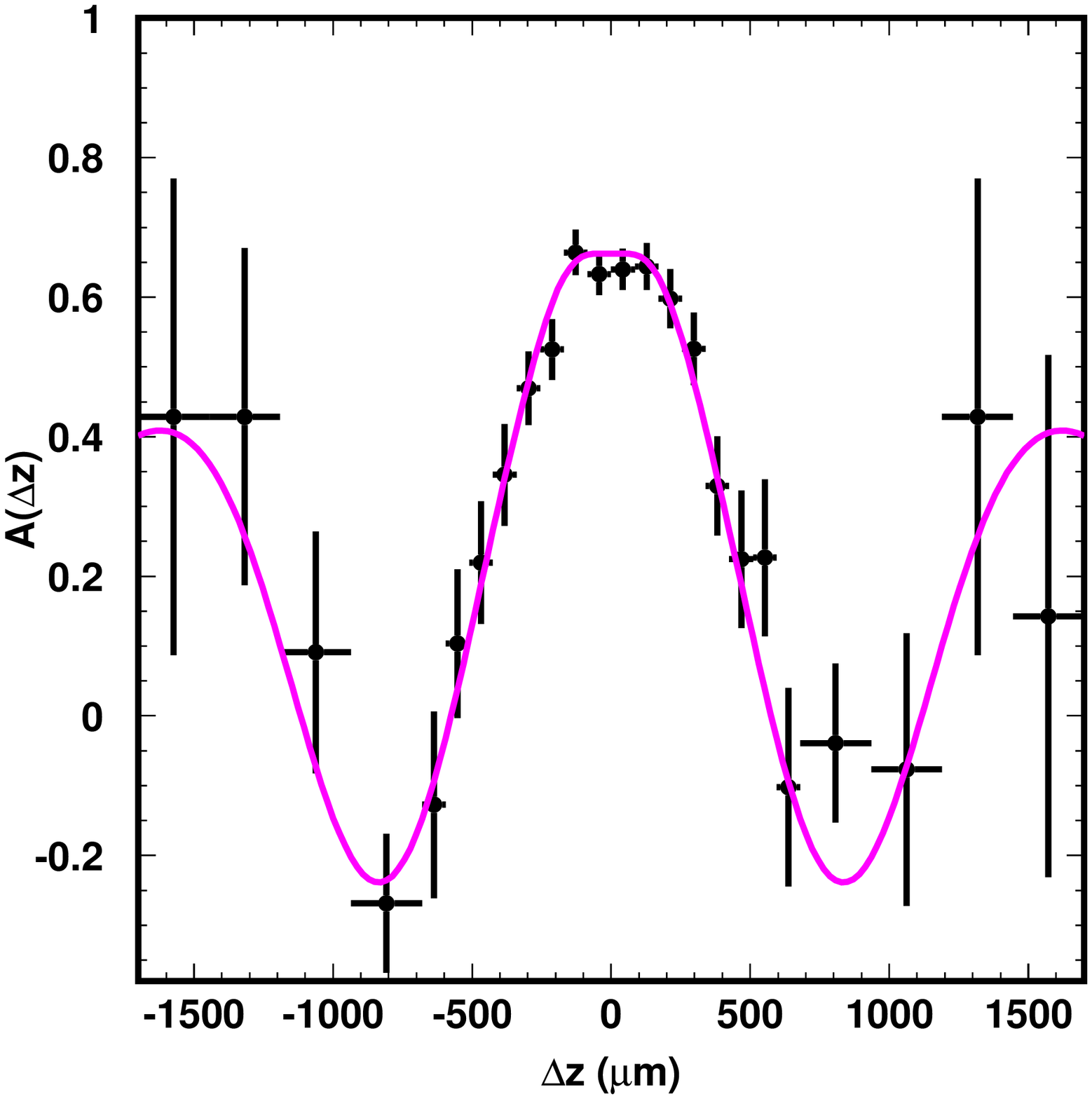}
}}
 \caption{\it
      Belle $\Delta m_d$ measurements based on 32 million $B \bar B$. 
      From left to right, dileptons, semileptonic decays, hadronic decays, and
 partially reconstructed $D^* \pi$ decays. 
    \label{belledm} }
\end{figure}

\begin{figure}[thbp]
  \centerline{\hbox{ \hspace{0.2cm}
    \includegraphics[width=3.5cm]{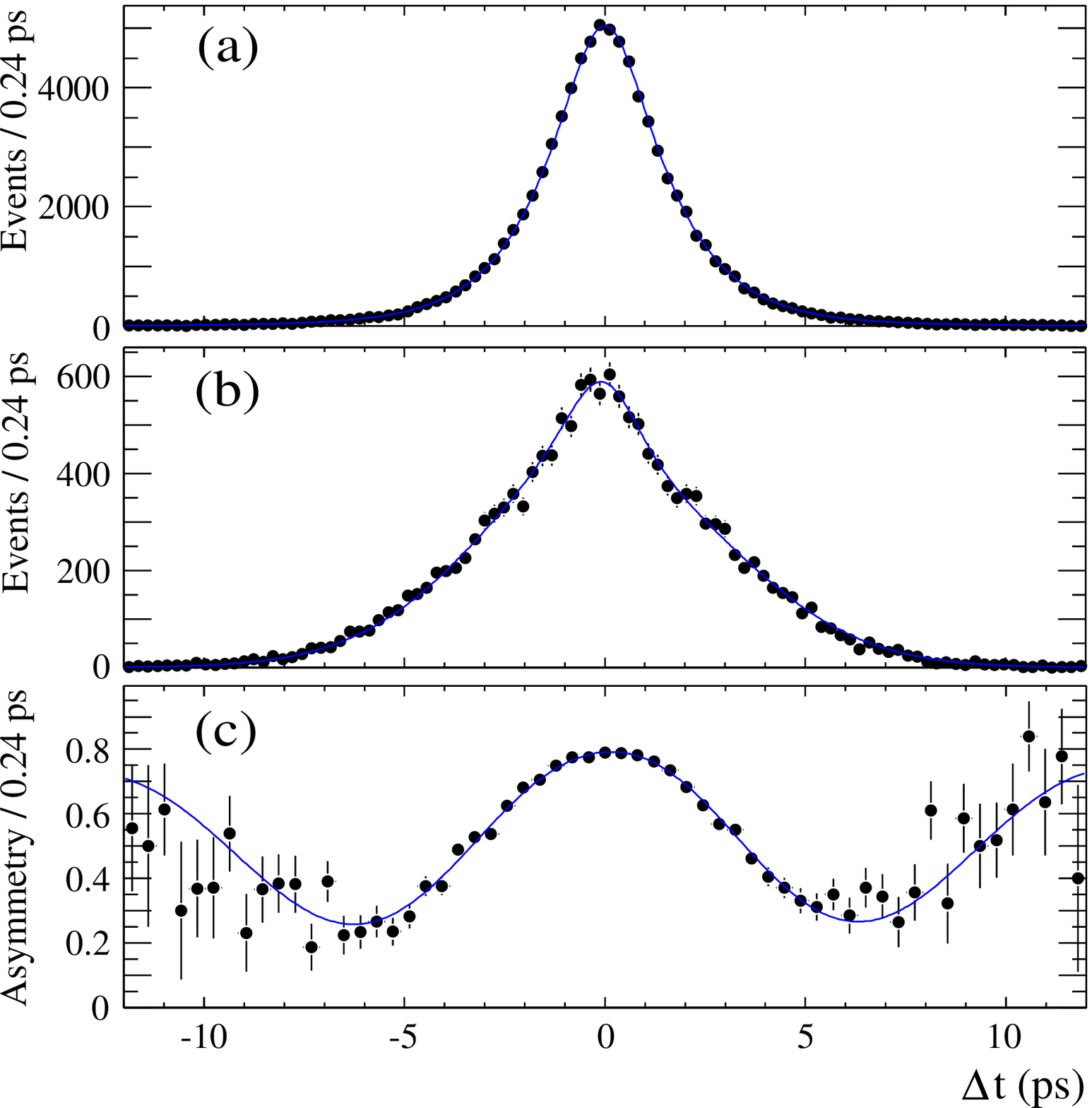}
   \includegraphics[width=3.5cm]{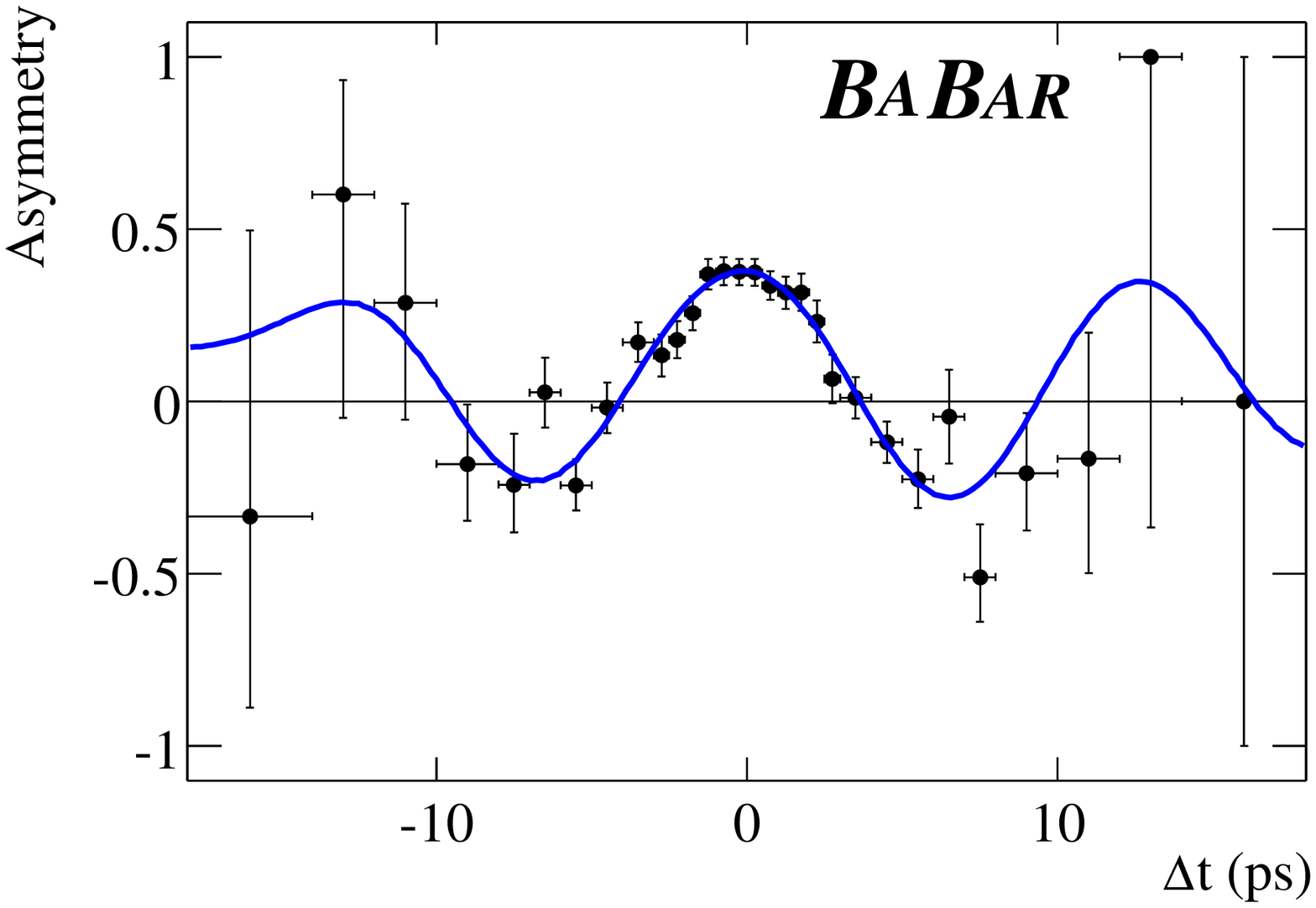}
   \includegraphics[width=3.5cm]{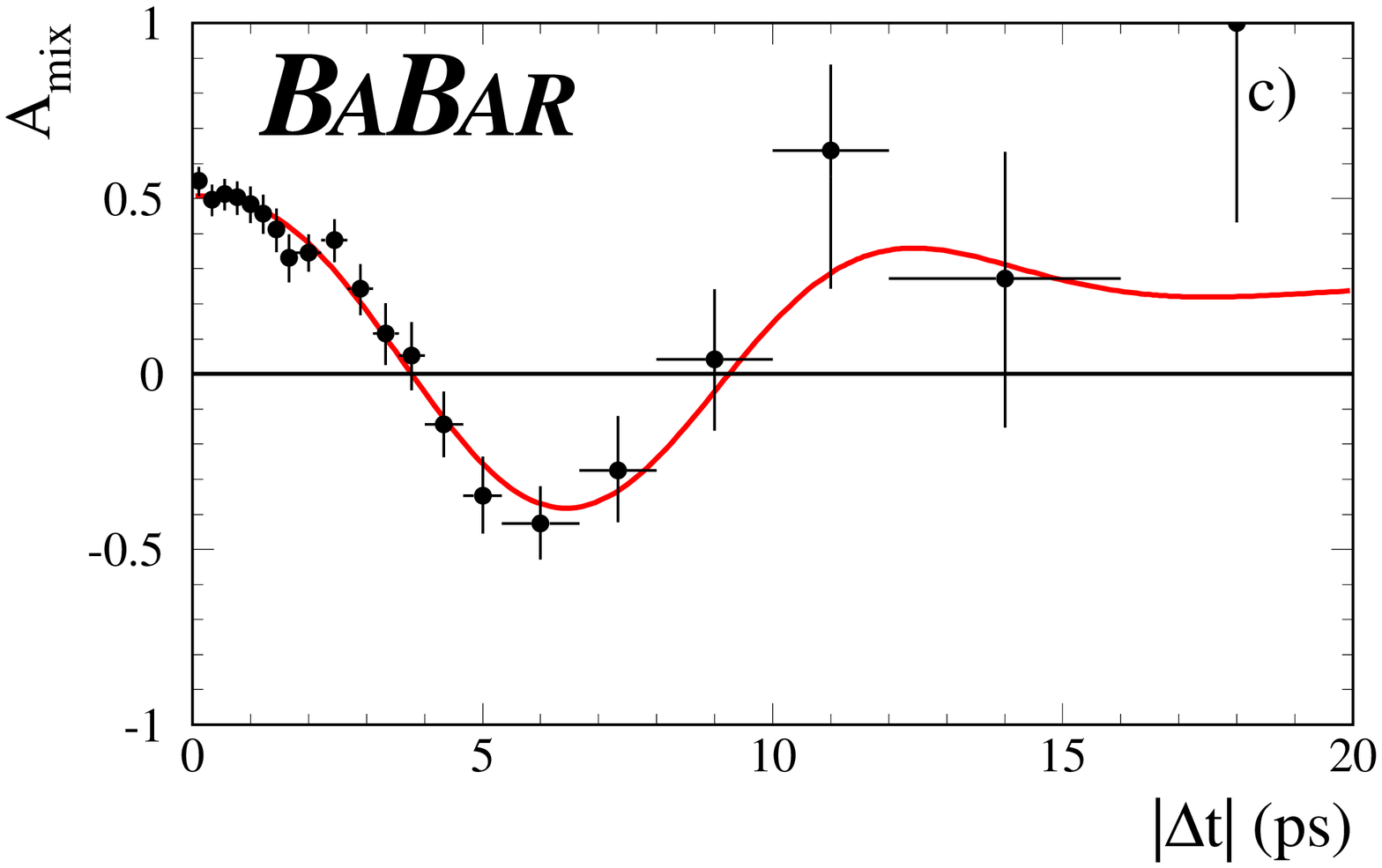}
    }
  }
 \caption{\it
       BaBar $\Delta m_d$ measurements. 
      From left to right, dileptons (23M $B \bar B$), 
        semileptonic decays (23M $B \bar B$), 
       hadronic decays (32M $B \bar B$). 
    \label{babardm} }
\end{figure}

\begin{figure}[htbp]
  \centerline{\hbox{ \hspace{0.2cm}
    \includegraphics[width=10cm]{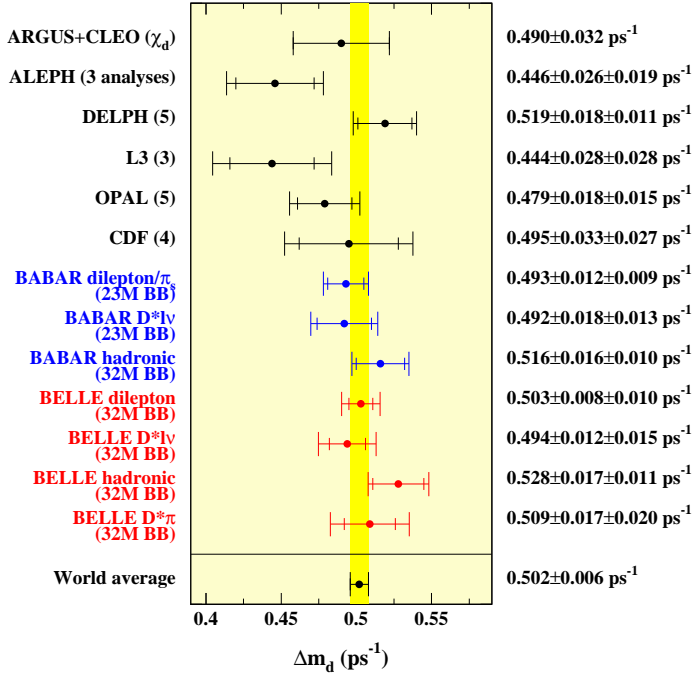}
    }
  }
 \caption{\it
      Present status of $\Delta m_d$ measurements.
    \label{dm-summary} }
\end{figure}

Mass and flavor eigenstates of the neutral $B$ meson states are
expressed  by
\begin{equation}
\mid B_1> = p\mid B^0> + q\mid \overline{B^0}>, ~~~
\mid B_2> = p\mid B^0> - q\mid \overline{B^0}>.
\end{equation}
Well defined time dependence of ($B_1$, $B_2$) and flavor-specific
decays of ($B^0$, $\bar{B^0}$) lead to the $B^0 \bar{B^0}$ oscillation. 
Probabilities of observing the two $B$ mesons as having the
opposite-flavor (OF) or having the same-flavor (SF) at $\Delta t$ 
are expressed by
\begin{equation}
P^{\mathrm{OF}} \propto  
                 \frac{e^{-|\Delta t/\tau_{B^0}|}}{4\tau_{B^0}} 
                           \left[ 1 + \cos(\Delta m_d \Delta t) \right],~~~ 
P^{\mathrm{SF}} \propto 
                 \frac{e^{-|\Delta t/\tau_{B^0}|}}{4\tau_{B^0}} 
                          \left[ 1 - \cos(\Delta m_d \Delta t) \right].
\end{equation}
The mixing parameters can be obtained either by reconstructing 
one $B$ in flavor-specific modes such as $D^{(*)} \pi$, $D^{(*)} \rho$,
$D^{(*)} \ell \nu$, and flavor-tagging the other $B$ using information
of remaining tracks in the event, or by using dilepton events. 
For the $\sin 2\phi_1$ measurement, we reconstruct one $B$ as $CP$
eigenstates such as $J/\psi K_S$.  
The OF-SF asymmetries that were measured by
Belle~\cite{belle-dm} and BaBar~\cite{babar-dm} are
shown in Fig.~\ref{belledm} and ~\ref{babardm}. 
The results are summarized in Figure~\ref{dm-summary}. 
A combined result of BaBar and Belle is 
$\Delta m_d = 0.504 \pm 0.007~\mathrm{ps^{-1}}$ which dominates the
world average of $\Delta m_d = 0.502 \pm 0.006~\mathrm{ps^{-1}}$. 

\section{ $B \bar B$ mixing in Standard Model}
In the Standard Model, box-diagram is responsible for 
$B \bar{B}$ mixing, and expressed as 
$\Delta m_d = m_{\mathrm{H}}-m_{\mathrm{L}}=2|M_{12}|$ 
where
\begin{equation}  
M_{12} = -\frac{G_F^2 m_W^2 \eta_B m_B B_B f_B^2} {12 \pi^2}
         S_0 (m_t^2/m_W^2) (V_{td}^* V_{tb})^2.
\end{equation}
Here $B_1$ and $B_2$ are redefined as $B_{\mathrm{H}}$ and
$B_{\mathrm{L}}$. Extraction of  $|V_{td}|$ from $\Delta
m_d$ is dominated by a large uncertainty in 
$f_{B_d} \sqrt{B_{B_d}} = 230 \pm 40$~MeV~\cite{PDG}. Improved 
lattice QCD calculations and $\Delta m_s$ measurements are waited.

The mixing also has an absorptive part 
$\Delta \Gamma = \Gamma_{\mathrm L} - \Gamma_{\mathrm H} =
2|\Gamma_{12}|$, which is tiny in the Standard Model.  
\begin{equation} 
\left| \frac{\Gamma_{12}}{M_{12}} \right| \sim 
          \frac{\Delta \Gamma}{\Gamma} 
            \simeq \frac{3\pi}{2}\frac{m_b^2}{m_W^2}
         \frac{1}{S_0 (m_t^2/m_W^2)} 
        \sim 5\times 10^{-3} (\pm 30\%).
\end{equation}
Any deviation will be difficult to explain in the Standard
Model, which of course makes this measurement very interesting. 
For non-zero $\Delta \Gamma$, the time-dependent decay rates
for the flavor-specific state ($B \to f (\bar f)$) must be modified  as 
\begin{equation}
\left[1 \pm \cos(\Delta m_d \Delta t)\right] \to
 \left[ \cosh \frac{\Delta \Gamma_d \Delta t}{2} \pm \cos (\Delta m \Delta t) 
      \right] 
\end{equation}
while for $CP$ eigenstate ($B^0 \to f_{CP}$, $CP$-even ($CP$-odd)), it must
be modified  as 
\begin{equation}
\left[1 \pm\sin 2\phi_1\sin(\Delta m_d \Delta t)\right]  
\to    \left[ \cosh \frac{\Delta \Gamma_d \Delta t}{2} 
           \mp \cos 2\phi_1 \sinh \frac{\Delta \Gamma_d \Delta t}{2} 
           \pm \sin 2\phi_1 \sin (\Delta m \Delta t) \right].
    \end{equation}

$CP$ violation in the $B \bar B$ mixing leads to $|q/p| \neq 1$
and it is related to $\Gamma_{12}$ and $M_{12}$ as
\begin{equation}
1 - |\frac{q}{p}|^2 \simeq Im\left(\frac{\Gamma_{12}}{M_{12}}\right). 
\end{equation}
In the Standard Model, $|q/p|$ is less than $10^{-3}$ 
because $|\Gamma_{12}/M_{12}| \sim 5\times 10^{-3}$ and  
$\phi_{M_{12}} - \phi_{\Gamma_{12}} = \pi + O(m_c^2/m_b^2)$.  
Probabilities of observing the SF events are given for ++ and --
combinations separately by 
$P^{\mathrm{SF}}_{++} = |p/q|^2 \cdot P^{\mathrm{SF}}$ and 
             $P^{\mathrm{SF}}_{--} = |q/p|^2 \cdot P^{\mathrm{SF}}$.  
Thus a charge asymmetry in the SF events appears if $CP$ is violated. 

$CPT$ violation leads to $p \neq p^{\prime}$ and/or $q \neq
q^{\prime}$ where the  $B$ meson states are described by 
$ \mid B_H> = p\mid B^0> + q\mid \overline{B^0}>,~~~ 
 \mid B_L> = p^{\prime}\mid B^0> - q^{\prime}\mid \overline{B^0}>$.
We introduce variables $\theta$ and $\phi$ 
where $ q/p = \tan (\frac{\theta}{2}) e^{i\phi}$, and 
$ q^{\prime}/p^{\prime} = \cot (\frac{\theta}{2}) e^{i\phi}$. 
The time dependence of the OF decay is modified as 
\begin{equation}
1 + \cos (\Delta m_d \Delta t)  \to  
[ 1 + |\cos \theta|^2 + ( 1 - |\cos \theta|^2) 
                                              \cos(\Delta m_d \Delta t)
         - 2 Im(\cos \theta) \sin(\Delta m_d \Delta t) ]. 
\end{equation}
A time-dependent asymmetry in the OF events can appear if $CPT$ is violated~
\cite{mohapatra}. 

\section{ Results of $\Delta \Gamma/\Gamma$, $|q/p|$, $\cos \theta$}
BaBar has performed a global fit to the fully reconstructed hadronic
events from the 88M $B \bar B$ sample and extracted $\Delta
\Gamma/\Gamma$, $|q/p|$, $Re(\cos \theta)$, and $Im (\cos \theta)$~\cite{babar-dg}. 
BaBar also obtained $|q/p|$ from the dilepton events in the 23M $B
\bar B$ sample~\cite{babar-dl}. Belle determined $Im (\cos \theta)$ and $Re (\cos
\theta)$ using the dilepton events in the 32M $B \bar B$
sample~\cite{belle-dm}.  
Results are summarized in Table~\ref{cpt}.

\begin{table}[bhtp]
\centering
\caption{ \it Results of $\Delta \Gamma/\Gamma$, $|q/p|$, $\cos \theta$. 
The parameter $z$ is equivalent to $\cos \theta$. 
$\mathrm{sgn}(\mathrm{Re}\lambda_{CP})=+1$ in SM. 
$\mathrm{Re} \lambda_{CP}/|\lambda_{CP}| \simeq 0.672 \pm 0.068$.}
\vskip 0.1 in
\begin{tabular}{|l|l|c|c|} \hline
 & data & variables & result \\
\hline
\hline
BaBar & hadronic &
      $\mathrm{sgn}(\mathrm{Re} \lambda_{CP})\Delta \Gamma/\Gamma$ & 
        $-0.008 \pm 0.037 \pm 0.018$  \\
 & &$|q/p|$ & $1.029 \pm 0.013 \pm 0.011$   \\
 & &$\mathrm{Re} \lambda_{CP}/|\lambda_{CP}|\mathrm{Re}z$ & 
    $0.014 \pm 0.035 \pm 0.034$ \\
 & &$\mathrm{Im}z$ & $0.038 \pm 0.029 \pm 0.025$   \\
BaBar &dileptons &$|q/p|$ & $0.998 \pm 0.005 \pm 0.007$  \\
Belle &  dileptons &$\mathrm{Im}(\cos \theta)$ & $0.03 \pm 0.01 \pm 0.03$ \\ 
      &            &$\mathrm{Re}(\cos \theta)$ & $ 0.00 \pm 0.12 \pm 0.01$ \\
\hline
\end{tabular}
\label{cpt}
\end{table}

\section{ $\sin 2\phi_1$ from $J/\psi K_S$ and other 
                              $b \to c \bar{c} s$ decays}
Asymmetry of time-dependent decay rates between ($B^0 \to f$) and 
($\bar{B^0} \to \bar{f}$) for the final state 
                    $f = \bar{f} = f_{CP}$ is expressed by
\begin{equation}
a_f (t) = \frac{\Gamma (\bar{B^0}(t) \to f) - 
                          \Gamma (B^0 (t) \to f)}
                {\Gamma (\bar{B^0}(t) \to f) + 
                         \Gamma (B^0 (t) \to f)}
 = \frac{2 Im \lambda_f}{ |\lambda_f|^2+1} \sin (\Delta m t) + 
   \frac{|\lambda_f|^2-1}{ |\lambda_f|^2+1} \cos (\Delta m t).
\end{equation}
Information of $CP$ violation is in a quantity $\lambda_f$. Namely 
$Im \lambda_f \neq 0 $ results in mixing-assisted $CP$ violation, and 
$|\lambda_f| \neq 1$ results in direct $CP$ violation. 
The $\lambda_f$ is defined as  
$\lambda_f = (q/p) \times <f|H|\bar{B^0}>/<f|H|B^0>$  
where
the $B \bar B$ mixing contribution is given by $q/p = 
(V^*_{tb} V_{td})/(V_{tb} V^*_{td})$ which is equal to 
$e^{-2i\phi_1}$ in the Standard Model.  

For the $J/\psi K_S$ final state (Fig.~\ref{jpsiks} followed by $K^0 \to
K_S$), $\lambda$  is given by
\begin{equation}
   \lambda (J/\psi K_S)
           =  \frac{V^*_{tb} V_{td}}{V_{tb} V^*_{td}} \cdot
   \eta_{J\psi K_S} \cdot
     (\frac{V_{cb}V^*_{cs}}{V^*_{cb}V_{cs}}) \cdot
     (\frac{V^*_{cd}V_{cs}}{V^*_{cs}V_{cd}}). 
\end{equation}
Here $\eta_f$ is $CP$ eigenvalue of the $f$ state. We obtain 
$Im \lambda(J/\psi K_S) = \sin 2\phi_1$ and 
$Im \lambda(J/\psi K_L) = -\sin 2\phi_1$. 

\begin{figure}[htbp]
  \centerline{\hbox{ \hspace{0.2cm}
    \includegraphics[width=4cm]{psiks.epsi}
    }
  }
 \caption{\it
      Diagram for $B^0 \to J/\psi K_S$.
    \label{jpsiks} }
\end{figure}

\begin{figure}[bhtp]
\begin{minipage}{6.7cm}
  \centerline{\hbox{ \hspace{0.2cm}
    \includegraphics[width=6.5cm]{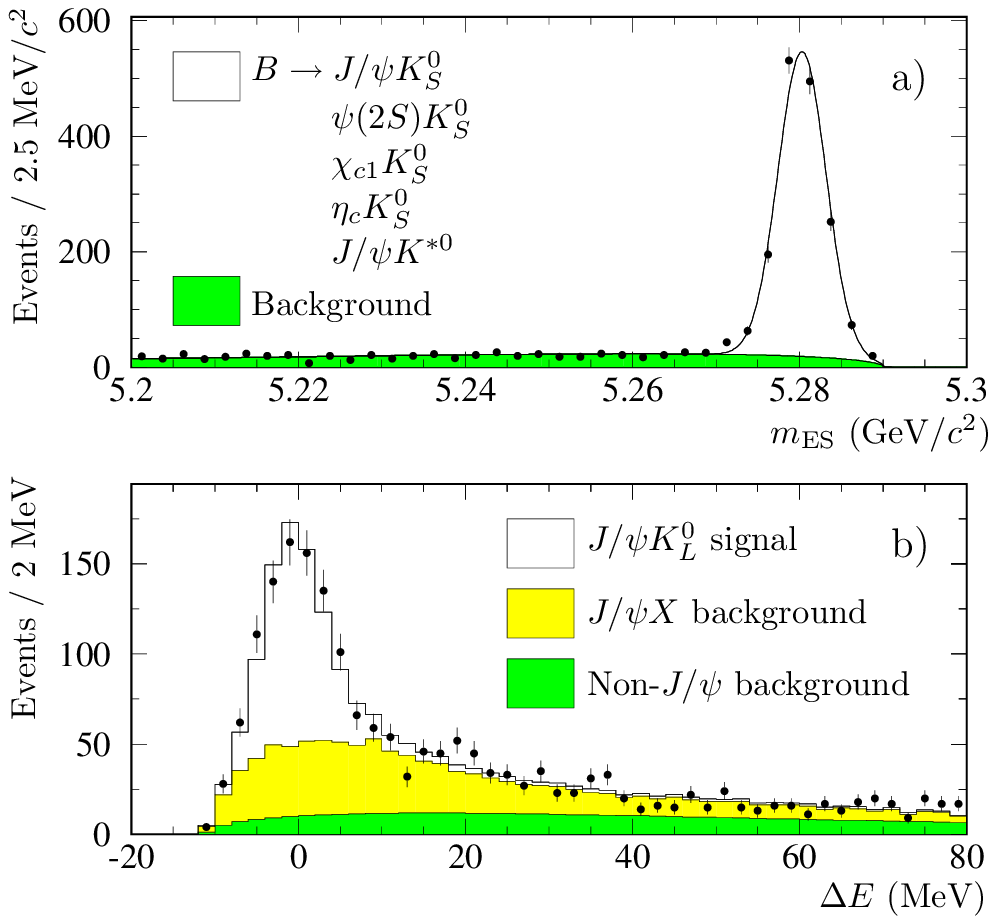}
    }}
\end{minipage}
\begin{minipage}{8cm}
  \centerline{\hbox{ \hspace{0.2cm}
    \includegraphics[width=4cm]{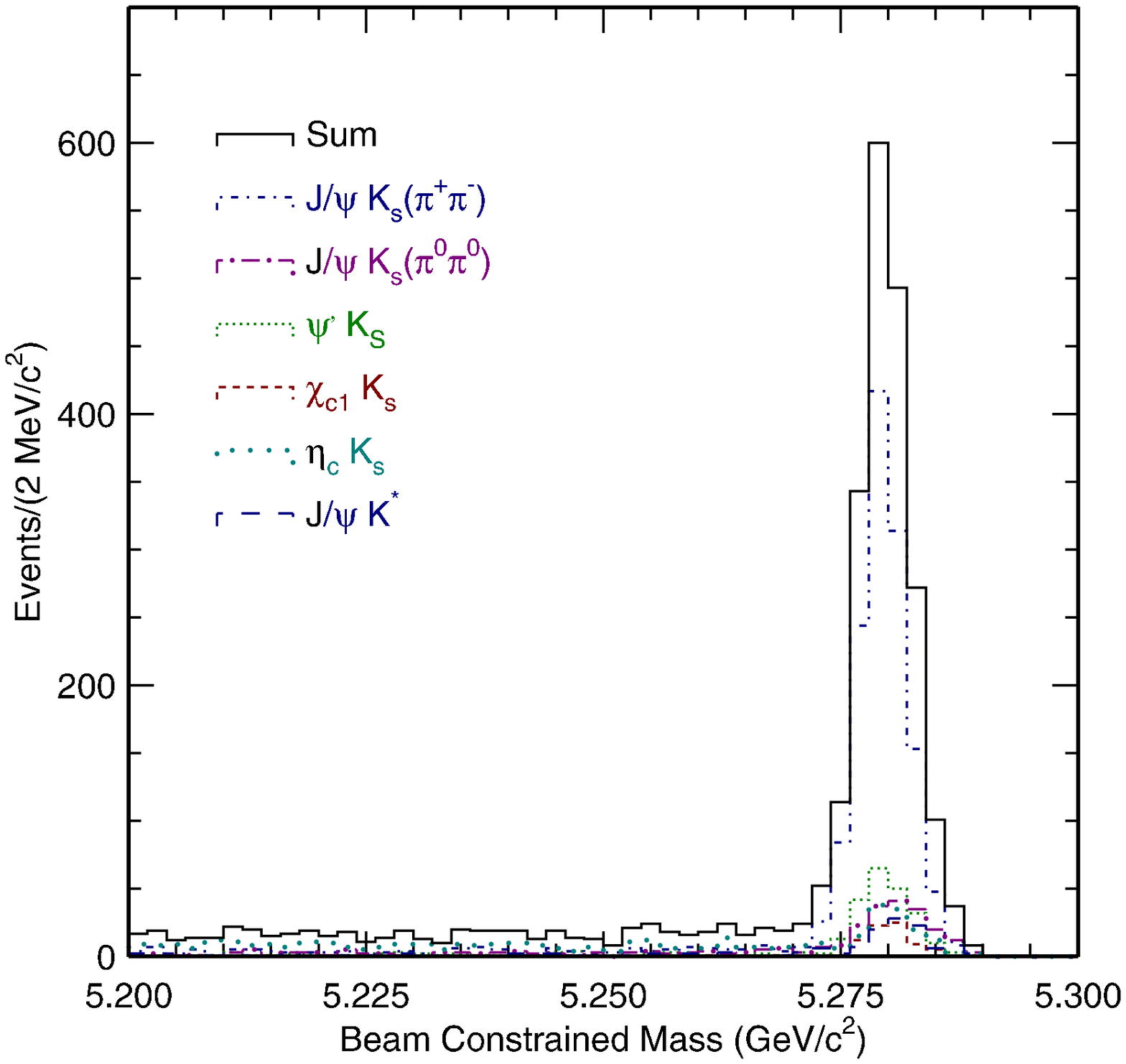}
    \includegraphics[width=4cm]{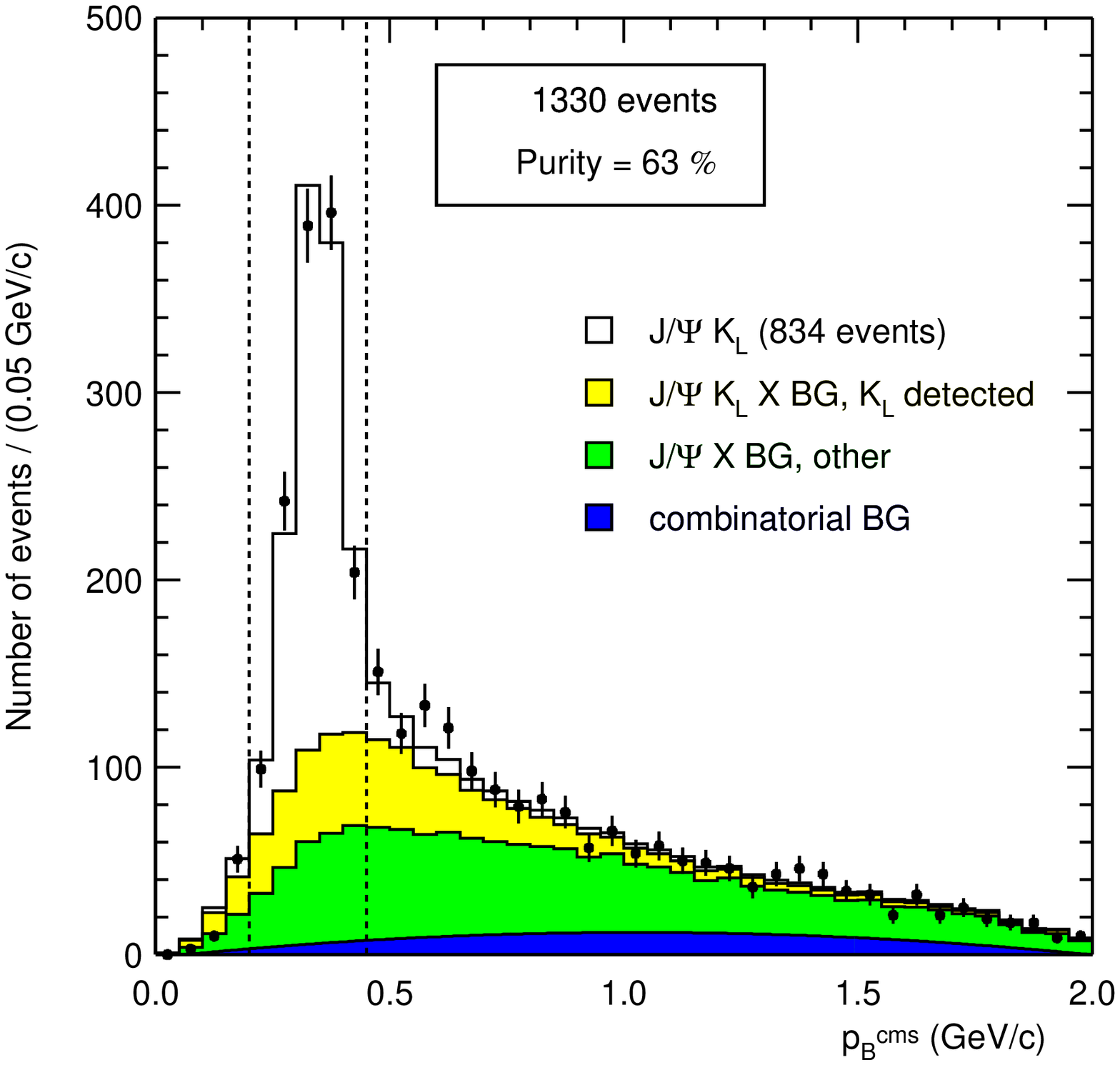}
    }}
\end{minipage}
 \caption{\it
     (Left) Beam-energy substituted mass distribution for the $\eta_{CP} =-1$ final
 states and $\Delta E$ distribution for the $J/\psi K_L$ final state for
 BaBar. (Right)
      Beam-energy substituted mass distribution for  the $\eta_{CP} =-1$
 final states and $p^*_{B}$ distribution for the $J/\psi K_L$ final
 state for Belle.
    \label{event} }
\end{figure}

Methods for the event selections are given in detail in
references~\cite{babar-psiks} and ~\cite{belle-psiks}.
The results presented here are based on the data set of 88M $B \bar B$ 
for BaBar and 85M $B \bar B$ for Belle. Both group used 
$J/\psi K_S$, $\psi^{\prime} K_S$, $\chi_{c1} K_S$, $\eta_c K_S$,
$J/\psi K^*$, and $J/\psi K_L$ final states. 
Except for the $J/\psi K_L$ final state, the candidate events peak in the 
mass distributions for reconstructed $B$ mesons. For the $J/\psi K_L$ events,
two-body decay of $B$ must be assumed since the $K_L$ energy cannot be
detected. BaBar uses the energy-difference, $\Delta E$, between
reconstructed $B$ and beam energy, whereas Belle uses the center-of-mass
momentum of reconstructed $B$, $p_B^*$. They are shown in  
Fig.~\ref{event}. 

\begin{figure}[bhtp]
  \centerline{\hbox{ \hspace{0.2cm}
    \includegraphics[width=3.5cm]{dt-asym-golden-v3-bgblack.epsi}
    \includegraphics[width=3.5cm]{dt-asym-klong-v3-bgblack.epsi}
    \includegraphics[width=3.5cm]{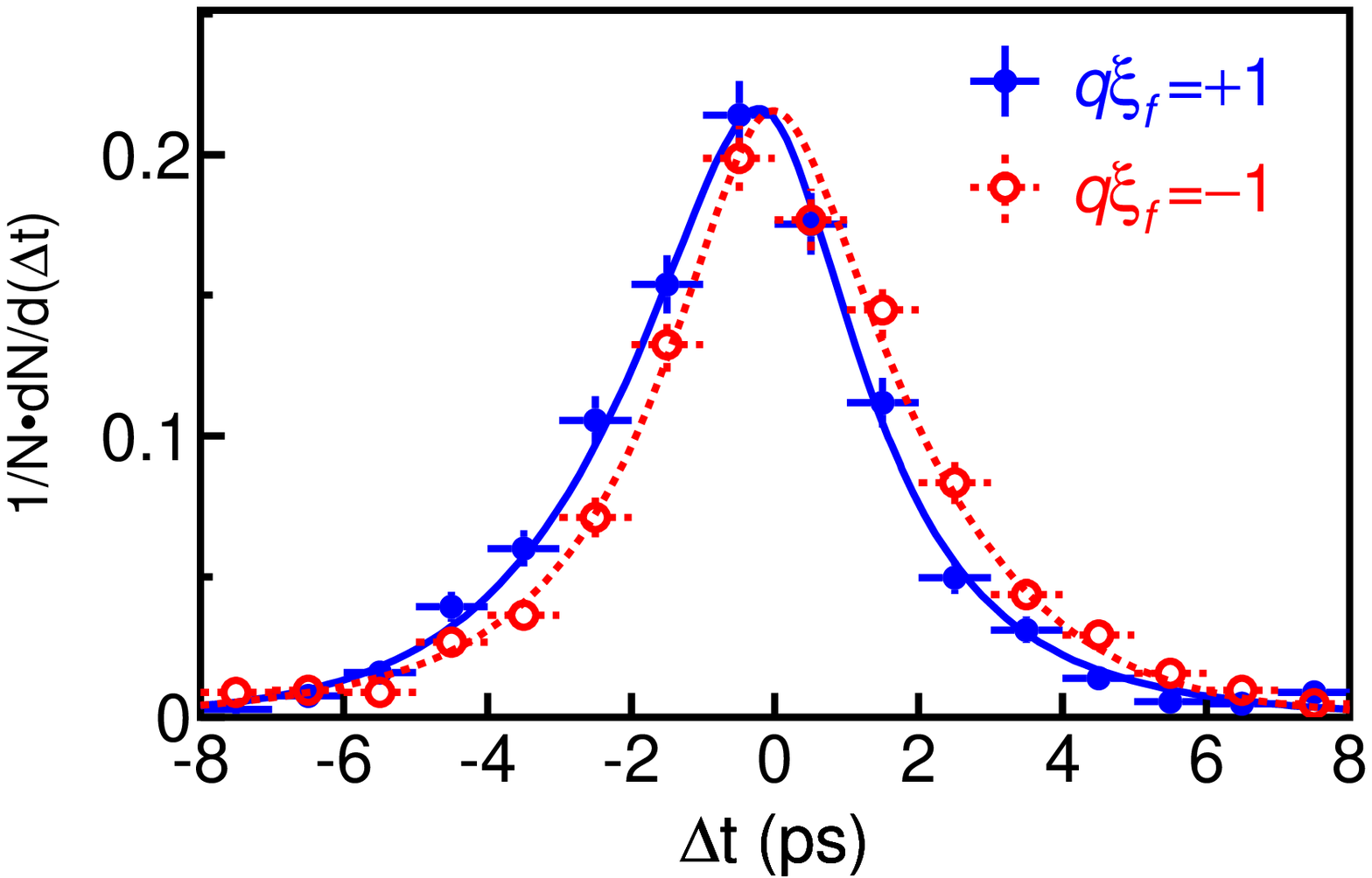}
    \includegraphics[width=3.5cm]{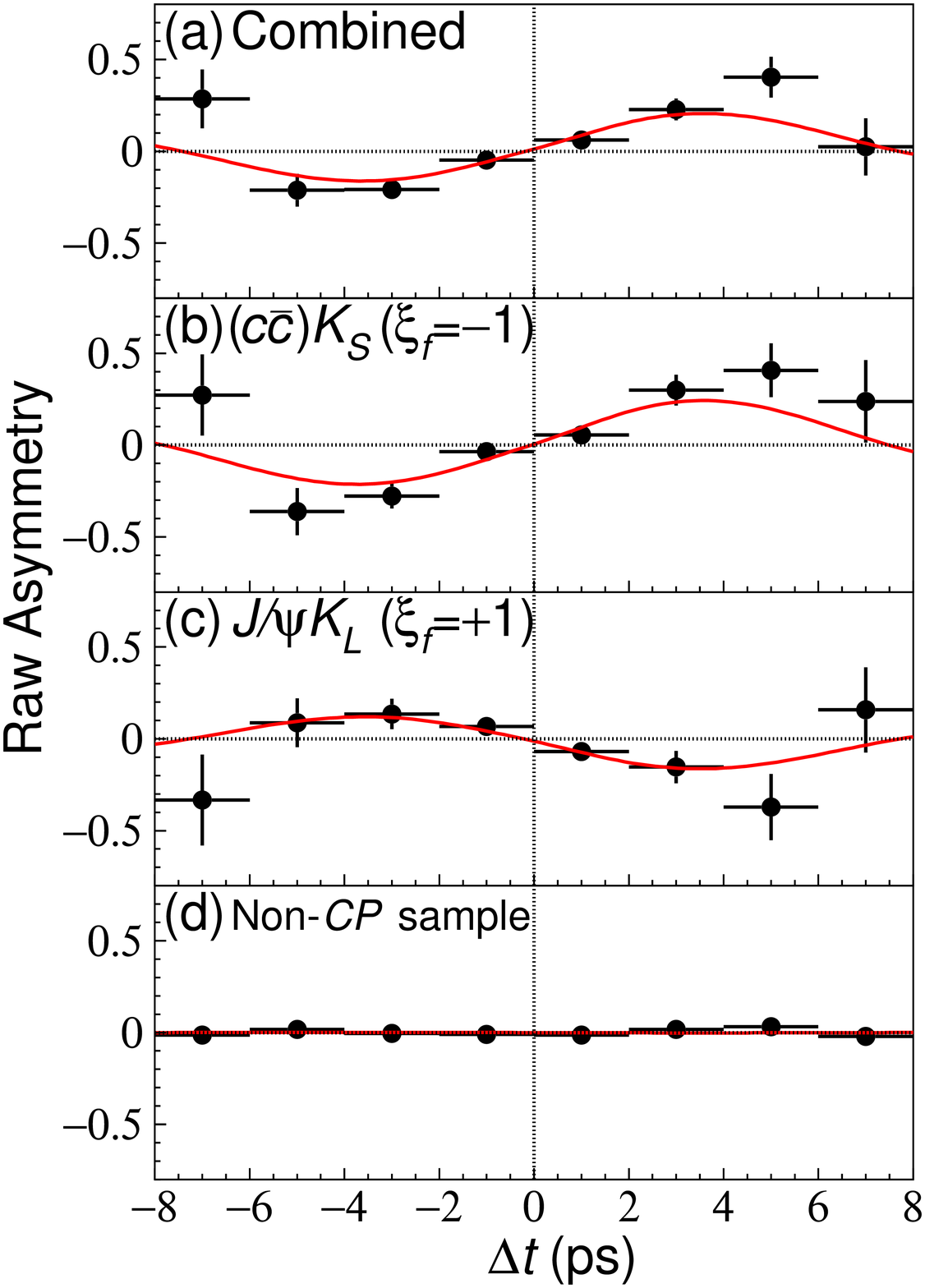}
    }
  }
 \caption{\it 
      BaBar $\Delta t$ distributions and asymmetries for $CP$-odd final
 states (far-left)  and $J/\psi K_L$ state (2nd-left).
      Belle $\Delta t$ distributions for a sum of $B^0$-tagged $J/\psi K_L$ and 
$\overline{B^0}$-tagged $CP$-odd states (labeled as $q\xi_f =+1$) and
 for a sum of  $\overline{B^0}$-tagged $J/\psi K_L$ and 
$B^0$-tagged $CP$-odd states (labeled as $q\xi_f =-1$)
 (2nd-right). Far-right are Belle asymmetries for  $q\xi_f =+1$ and
 $q\xi_f =-+1$ samples combined (a), each separately (b) and (c), 
and for non-$CP$ sample (d). 
    \label{dt} }
\end{figure}

Extraction of $\sin 2\phi_1$ from the $\Delta t$ distributions are done
by maximize  a likelihood $L = \prod_{i}P_{i}$ (i $\cdot \cdot \cdot$each candidate
 event). The probability of each candidate event is described by
\begin{equation}
P_i = \int \left[f_{\rm sig} 
        P_{\rm sig}(\Delta t^{\prime})
      R_{\rm sig}(\Delta t -\Delta t^{\prime}) + 
      (1-f_{\rm sig})P_{\rm bkg}(\Delta t^{\prime})
           R_{\rm bkg}(\Delta t -\Delta t^{\prime})\right] 
          d\Delta t^{\prime}
\end{equation}
where $f_{\rm sig}$ is signal fraction of candidate event,  
$P_{\rm sig}$ and $P_{\rm bkg}$ are the probability density functions,
 and $R_{\rm sig}$ and $R_{\rm bkg}$ are the  $\Delta t$ resolutions. 
The $\Delta t$ distributions and asymmetries are
shown in Fig.~\ref{dt} together with their fit results.  

\begin{figure}[htbp]
  \centerline{\hbox{ \hspace{0.2cm}
    \includegraphics[width=6.5cm]{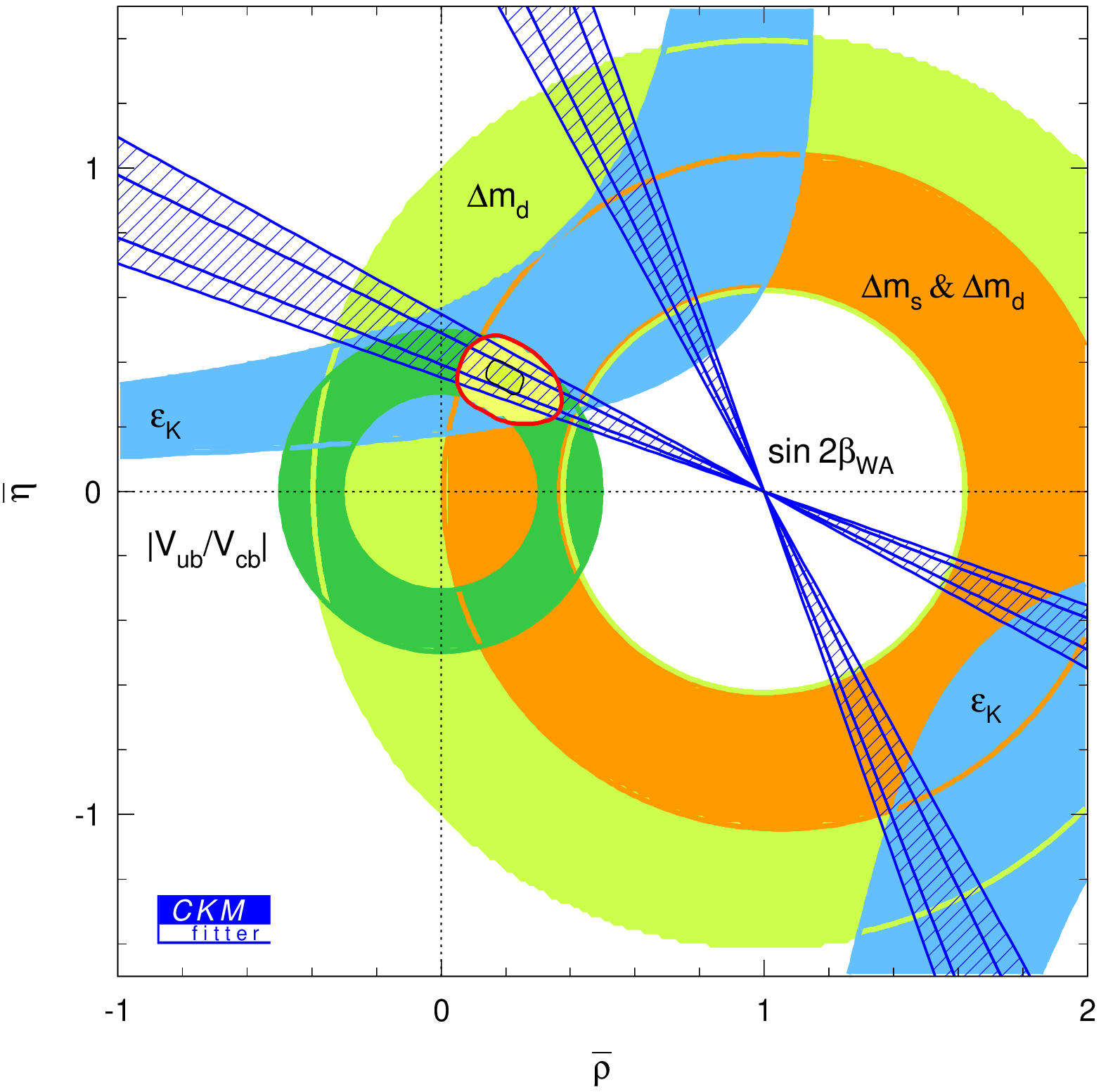}
    }
  }
 \caption{\it
      Shaded area are for 1$\sigma$ and 2$\sigma$ regions from the
 BaBar-Belle combined value of $\sin 2\phi_1$.  90\% (5\%) CL contours
 from a global CKM fit are also shown. 
    \label{ckm} }
\end{figure}

The BaBar results are $\sin 2\phi_1 = 0.741 \pm 0.067 \pm 0.034$ and 
$|\lambda | = 0.948 \pm 0.051 \pm 0.030$, while the Belle results are 
$\sin 2\phi_1 = 0.719 \pm 0.074 \pm 0.035$ and 
$| \lambda | = 0.950 \pm 0.049 \pm 0.025$. 
A combined result is $\sin 2\phi_1 = 0.734 \pm 0.055$. 
Fig.~\ref{ckm} shows an allowed region of ($\rho$-$\eta$) plane from 
the $\sin2\phi_1$ measurement and from a global CKM fit without using
$\sin 2\phi_1$. Agreement is excellent.

\section{ $\sin 2\phi_1$ from  loop diagram decays}
\subsection{$\phi K_S$ }
\begin{figure}[htbp]
  \centerline{\hbox{ \hspace{0.2cm}
    \includegraphics[width=4cm]{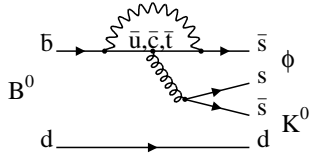}
    }
  }
 \caption{\it
      Standard Model contribution to $B^0 \to \phi K_S$. 
    \label{phiks-diagram} }
\end{figure}

The $B^0 \to \phi K_S$ decay has only $b \to ss\bar{s}$ penguin
contribution in the Standard Model (Fig.~\ref{phiks-diagram}). 
Leading term has a CKM factor of $V_{cb} V^*_{cs}(P_c - P_t) =A\lambda^2
(P_c - P_t)$, where $P_q$ are the penguin amplitudes. This is same as
the CKM factor for $B^0 \to J/\psi K_S$. Next-to-leading
term $V_{ub} V^*_{us}(P_u - P_t)=A\lambda^4 (\rho -i\eta) (P_u - P_t)$
has a different phase, but is suppressed by $\lambda^2 \simeq 5\%$. 
Since $\eta_{\phi K_S} = -1$, $\sin 2\phi_1$ measured in this mode
should be the same as that for the $J/\psi K_S$ in the Standard Model.
In order to allow room for new physics, we parameterize the asymmetry 
distribution by
\begin{equation}
 a_f (\Delta t) = S_f \sin (\Delta m_d \Delta t) + 
        A_f \cos (\Delta m_d \Delta t)
\end{equation}
where 
\begin{equation}
S_f=\frac{2 Im \lambda_f}{|\lambda_f|^2 + 1} 
         (\simeq -\eta_f \sin 2\phi_1~\mathrm{in~SM}), ~~~
A_f=-C_f=\frac{|\lambda_f|^2 - 1}{|\lambda_f|^2 + 1} (\simeq 0~\mathrm{in~SM}).
\end{equation}
Any deviation would be an indication of new physics in penguin loop. 

The BaBar results based on 84M $B \bar B$~\cite{babar-phiks} are 
$S_{\phi K_S} = -0.18 \pm 0.51 \pm 0.07$ and 
$A_{\phi K_S} = +0.80 \pm 0.38 \pm 0.12$, 
whereas the Belle results based on 85M $B \bar B$~\cite{belle-phiks} are 
$S_{\phi K_S} = -0.73 \pm 0.64 \pm 0.22$ and 
$A_{\phi K_S} = -0.56 \pm 0.41 \pm 0.16$.

\subsection{$\eta^{\prime} K_S$}
This mode is contributed by $b \to s s \bar{s}$ penguin, 
$b \to s d \bar{d}$ penguin,  and $b \to u$ tree diagrams (Fig.~\ref{etapks}). 
\begin{figure}[htbp]
  \centerline{\hbox{ \hspace{0.2cm}
    \includegraphics[width=4.0cm]{etapss.epsi}
    \includegraphics[width=4.0cm]{etapdd.epsi}
    \includegraphics[width=4.0cm]{etapk_tree.epsi}
    }
  }
 \caption{\it
      Standard Model contributions to $B^0 \to \eta^{\prime} K_S$. 
    \label{etapks} }
\end{figure}
In the Standard Model, presence of additional $b \to s d \bar{d}$
 penguin does not cause any change from the $\phi K_S$ case, and only difference is
 the  additional $b \to u$ tree diagram which is only 5\% effect. 
Since $\eta_{\eta^{\prime} K_S}=-1$, we expect to have $S_f \simeq \sin 2\phi_1$.

The BaBar results based on 88.9M $B \bar B$~\cite{babar-etapks} are
$S_{\eta^{\prime} K_S}=+0.02 \pm 0.34 \pm 0.03 b$ and 
$A_{\eta^{\prime} K_S}=-0.10 \pm 0.23 \pm 0.03$, 
whereas the Belle results based on 85M $B \bar{B}$~\cite{belle-phiks} are
$S_{\eta^{\prime} K_S}=+0.71 \pm 0.37 ^{+0.05}_{0.06}$ and 
$A_{\eta^{\prime} K_S}=+0.26 \pm 0.22 \pm 0.03$.

\subsection{$K^+ K^- K_S$}
\begin{figure}[bhtp]
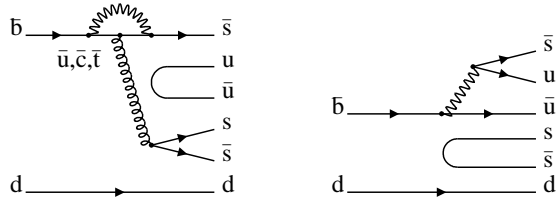

  \centerline{\hbox{ \hspace{0.2cm}
    \includegraphics[width=3cm]{kkks_p.epsi} \hspace{1cm}
    \includegraphics[width=3cm]{kkks_t.epsi}
    }
  }
 \caption{\it
      Standard Model contributions to $B^0 \to K^+ K^- K_S$. 
    \label{kkks} }
\end{figure}

This decay is contributed by $b \to s$ penguin and $b \to u$ tree
  diagrams (Fig.~\ref{kkks}). 
The Belle analysis for this decay mode shows that the
$b \to u$ tree contribution is negligible and furthermore $CP$
content of the final state is predominantly even ($\eta_{K^+ K^- K_S} =
+1$)~\cite{belle-phiks}. Therefore we expect  $S_f \simeq -\sin 2\phi_1$. 
The results based on 85M $B \bar B$ are 
$S_{K^+ K^- K_S}=-0.49 \pm 0.43 \pm 0.11$ and 
$A_{K^+ K^- K_S}=-0.40 \pm 0.33 \pm 0.10$.

\begin{figure}[tp]
  \centerline{\hbox{ \hspace{0.2cm}
    \includegraphics[width=10cm]{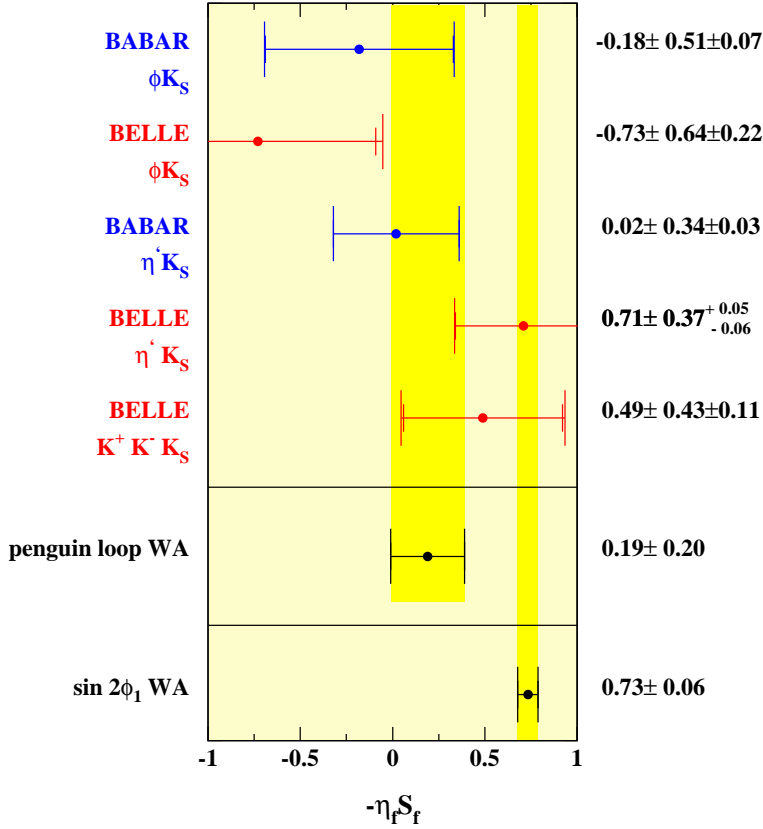}
    }
  }
 \caption{\it
      Summary of $ -\eta_f S_f$ measurements for
the penguin loop decays.
    \label{summary-loop} }
\end{figure}

Fig.~\ref{summary-loop} summarizes the ($ -\eta_f S_f$) measurements for
the penguin loop decays.
An average ``$\sin 2\phi_1$'' of those three penguin decays is $0.19 \pm
0.20$, about 2.5$\sigma$ off the Standard Model. 
We are entering an exciting era for exploring new physics through $\sin
2\phi_1$ measurements in different decay modes.

\section{ $\sin 2\phi_1$ from other modes}
\subsection{$J/\psi \pi^0$}
\begin{figure}[htbp]
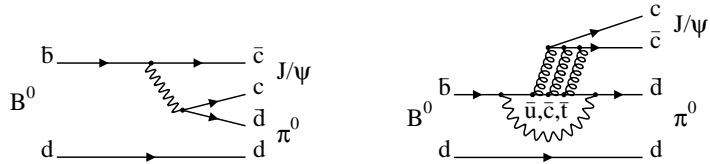

  \centerline{\hbox{ \hspace{0.2cm}
    \includegraphics[width=4cm]{psipi0.epsi} \hspace{1cm}
    \includegraphics[width=4cm]{psipi0_penguin.epsi}
    }
  }
 \caption{\it
      Standard Model contributions to $B ^0 \to J/\psi \pi^0$. 
    \label{psipi0} }
\end{figure}

In this mode, the tree and penguin contributions are of comparable
size(Fig.~\ref{psipi0}). The CKM factors are  
$V_{cb} V^*_{cd} = -A\lambda^3$ for the tree, and  
$V_{cb} V^*_{cd} (P_c -P_t)= -A\lambda^3 (P_c - P_t)$ and  
$V_{ub}V^*_{ud}(P_u -P_t) = A\lambda^3 (\rho - i\eta)(P_u -P_t)$ for the
penguins, respectively. 
In an extreme case of ignoring the penguin, we obtain $S_f \simeq -\sin
2\phi_1$ since $\eta_{J/\psi \pi^0}=+1$. If a deviation is seen,
presence of penguin should be suspected first. 
The BaBar results based on 88M $B \bar B$~\cite{babar-psipi0} are 
$S_{J/\psi \pi^0}=+0.05 \pm 0.49 \pm 0.16$ and 
$A_{J/\psi \pi^0}=-0.38 \pm 0.41 \pm 0.09$, 
whereas the Belle results based on 85M $B \bar B$~\cite{belle-psipi0} are
$S_{J/\psi \pi^0}=-0.93 \pm 0.49 \pm 0.08$ and 
$A_{J/\psi \pi^0}=-0.25 \pm 0.39 \pm 0.06$. 

\subsection{$D^{*+} D^{*-}$ and $D^{*+} D^-$ }
\begin{figure}[htbp]
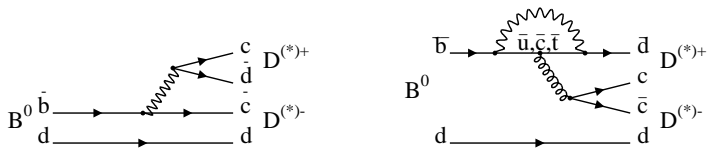

  \centerline{\hbox{ \hspace{0.2cm}
    \includegraphics[width=4cm]{ddtree.epsi} \hspace{1cm}
    \includegraphics[width=4cm]{ddpenguin.epsi}
    }
  }
 \caption{\it
      Standard Model contributions to $B^0 \to D^{(*)+} D^{(*)-}$.
    \label{dd} }
\end{figure}

These modes have similar ``penguin pollution'' as $J/\psi \pi^0$
(Fig.~\ref{dd}). The CKM factors are $V_{cb} V^*_{cd} = -A\lambda^3$ for
the tree, and
$V_{cb} V^*_{cd} (P_c -P_u) = -A\lambda^3 (P_c -P_u)$ and 
$V_{tb} V^*_{td}(P_t -P_u) = A\lambda^3 (1-\rho +i\eta) (P_t -P_u)$ for
the penguins, respectively. 
BaBar angular analysis~\cite{babar-dstdst} showed that $CP$ content of the 
$D^{*+} D^{*-}$ final state is predominantly even ($\eta_{D^{*+} D^{*-}}
\simeq +1$). In an extreme case of ignorin the penguin, we obtain  
$S_f \simeq -\sin 2\phi_1$. 
The $D^{*+} D^-$ final state is not a CP eigenstate. In an extreme case
of ignoring the penguin, we obtain $S_f^{\pm} = S_f^{\mp} \simeq -\sin 2\phi_1$.
The BaBar results based on 88M $B \bar B$~\cite{babar-ddst} are
\begin{eqnarray}
S_{D^* D}^{\pm}&=&-0.24 \pm 0.69 \pm 0.12,~~~
S_{D^* D}^{\mp}=-0.82 \pm 0.75 \pm 0.14 \nonumber \\
A_{D^* D}^{\pm}&=&+0.22 \pm 0.37 \pm 0.10,~~~
A_{D^* D}^{\mp}=+0.47 \pm 0.40 \pm 0.12.
\end{eqnarray}

\section{Summary}
Precision of $\Delta m_d$ has reached 1.2\%.
Attempt for observing higher order effect and possible new physics
effects in $B \bar B$ mixing are vigorously explored.
The $\Delta m_d$ measurements are an important testing ground for the 
$\Delta t$ measurement and flavor-tagging.
Precision of $\sin 2\phi_1$ has reached 8\%. Statistical error still dominates.
It is in good agreement with a global CKM fit (without $\sin 2\phi_1$). 
$|\lambda |$ is consistent with 1 in $b \to c \bar{c} s$ decays
as expected in the Standard Model.
New physics search by ``$\sin 2\phi_1$'' measurements in penguin
loops is well under way. 
``$\sin 2\phi_1$'' measurements for  ``penguin polluted'' decays
     were also pushed to find useful information.

\end{document}